\documentclass[traditabstract]{aa} 
\usepackage{amsmath}
\usepackage{graphicx}
\usepackage{txfonts}
\usepackage{comment}
\usepackage{natbib}
\bibpunct{(}{)}{;}{a}{}{,}
\usepackage{float}
\usepackage{calc}
\usepackage{textcomp}
\usepackage{placeins}
\usepackage{color}
\usepackage{rotating} 
\usepackage{lscape}
\usepackage{url}
\usepackage[colorlinks=true,linkcolor=blue,citecolor=blue]{hyperref}
\usepackage{subfig}
\usepackage[all]{draftcopy}
\usepackage{longtable}
\usepackage{array}
\usepackage{threeparttable}
\usepackage{lscape}
\usepackage{enumerate}
\DeclareTextSymbol{\degre}{OT1}{23}
\newcounter{savedfootnote}

\def \cigale{{\texttt{CIGALE}}}
\renewcommand{\epsilon}{\varepsilon} 

\begin{document}
\title{GOODS-ALMA 2.0: Last gigayear star formation histories of the so-called starbursts within the main sequence}

\author{
L.~Ciesla\inst{1},
C.~G\'omez-Guijarro\inst{2},
V.~Buat\inst{1,3}, 
D.~Elbaz\inst{2}, 
S.~Jin\inst{4,5,6}, 
M.~B\'ethermin\inst{1},
E.~Daddi\inst{2},
M.~Franco\inst{7},
H.~Inami\inst{8},
G.~Magdis\inst{4,5,6},
and 
B.~Magnelli\inst{2}
}

\institute{	
 Aix-Marseille  Universit\'e,  CNRS, LAM (Laboratoire d'Astrophysique de Marseille) UMR7326,  13388, Marseille, France
 \and
 Universit{\'e} Paris-Saclay, Universit{\'e} Paris Cit{\'e}, CEA, CNRS, AIM, 91191, Gif-sur-Yvette, France
  \and
 Institut Universitaire de France (IUF), Paris, France
 \and
 Cosmic Dawn Center (DAWN), Copenhagen, Denmark
 \and
 DTU-Space, Technical University of Denmark, Elektrovej 327, 2800, Kgs. Lyngby, Denmark
 \and
 Niels Bohr Institute, University of Copenhagen, Jagtvej 128, 2200, Copenhagen N, Denmark
 \and
 Department of Astronomy, The University of Texas at Austin, 2515 Speedway Blvd Stop C1400, Austin, TX 78712, USA
 \and
 Hiroshima Astrophysical Science Center, Hiroshima University, 1-3-1 Kagamiyama, Higashi-Hiroshima, Hiroshima 739-8526, Japan
}



   \date{Received; accepted}

  \abstract
{
Recently, a population of compact main sequence (MS) galaxies exhibiting starburst-like properties have been identified in the GOODS-ALMA blind survey at 1.1\,mm. 
Several evolution scenarios were proposed to explain their particular physical properties (e.g., compact size, low gas content, short depletion time). 
In this work, we aim at studying the star formation history (SFH) of the GOODS-ALMA galaxies to understand if the so-called ``starburst (SB) in the MS'' galaxies exhibit a different star formation activity over the last Gyr compared to MS galaxies that could explain their specificity.
We use the \cigale\ SED modelling code to which we add non-parametric SFHs.
To compare quantitatively the recent SFH of the galaxies, we define a parameter, the star formation rate (SFR) gradient that provides the angle showing the direction that a galaxy has followed in the SFR vs stellar mass plane over a given period.
We show that ``SB in the MS'' have positive or weak negative gradients over the last 100, 300, and 1000\,Myr, at odds with a scenario where these galaxies would be transitioning from the SB region at the end of a strong starburst phase.
Normal GOODS-ALMA galaxies and ``SB in the MS'' have the same SFR gradients distributions meaning that they have similar recent SFH, despite their different properties (compactness, low depletion time).
The ``SBs in the MS'' manage to maintain a star-formation activity allowing them to stay within the MS.
This points toward a diversity of galaxies within a complex MS.

}


   \authorrunning{Ciesla et al.}
   \titlerunning{GOODS-ALMA 2.0: last Gyr SFH}

   \maketitle

\section{\label{intro}Introduction}

The discovery of a tight relation between the star formation activity of galaxies and stellar mass opened a new window in our understanding of galaxy evolution \citep{Noeske07_SFseq,Elbaz07}. 
One of the main consequences of this relation is that galaxies form the bulk of their stars through secular processes rather than violent episodes of star formation. 
This relation between the star formation rate (SFR) and stellar mass (M$_*$), the so-called galaxies main sequence (MS), has been now constrained up to z$\sim$5 \citep[e.g.,][]{Schreiber18,Khusanova21,Topping22} and its slope and normalisation varies with cosmic time \citep[e.g.,][]{Speagle14,Schreiber15,Leja21}. 
However, what is striking is that the its dispersion seems to be constant whatever the stellar mass range or redshift studied \citep[e.g.,][]{Guo13,Schreiber15} although \cite{Leja21} recently claim a wider MS at redshift larger than 2 and at the high mass end ($\sim$10$^{11}$\,M$_{\odot}$). 
Several studies have shown that there is a coherent variation of the physical properties of galaxies within the MS such the gas fraction \citep{Magdis12,Wuyts11,Salmi12,Saintonge22} suggesting that the scatter of the MS is the result of physical processes rather than uncertainties on the measurements. 
However, the path that galaxies follow on the SFR-M$_*$ diagram and the processes that drive these evolutionary paths are still debated. 

Studying the properties of star forming galaxies within the scatter of the MS is of paramount importance to understand the mechanisms governing smooth and self-regulated galaxy evolution.
Recent studies using the Atacama Large Millimeter/submillimeter array (ALMA) have highlighted the existence of a population of starburst galaxies (SB) within the scatter of the MS.
These sources have short depletion timescales and enhanced star formation surface densities, exhibiting compact star formation traced by submm/mm dust continuum emission \citep{Elbaz18} or by radio emission \citep{Jimenez19}, hence their name.
Constraining the evolutionary scenario of these populations of compact galaxies and ``SB in the MS" is still to be done.
Several studies have advocated for compaction events predicted by galaxy formation models in which extended star-forming galaxies in the MS can secularly evolve into compact star-forming galaxies in the MS by funneling gas to their central regions yielding the build up of their stellar cores \citep[e.g.,][]{Dekel13,Zolotov15,Tacchella16}.
Others have proposed a funneling of the gas to the center driven by a violent episode of star formation typical of gas-rich mergers \citep[e.g.,][]{MihosHernquist96,Hopkins06,Toft14,GomezGuijarro18,Puglisi21}. 
In this scenario, compact star-forming galaxies in the MS would then be in the end of their starbursting phase having consumed most of their gas and just passing through the MS on their way to quiescence \citep{Elbaz18,GomezGuijarro19,Puglisi21}.

Studying the galaxies detected in the blind GOODS-ALMA 2.0 survey \citep{GomezGuijarro22a}, \cite{GomezGuijarro22b} showed that the ``SB in the MS" are extreme cases where the dust continuum areas are the most compact ones, associated with the shortest depletion timescales, lowest gas fractions, and the highest dust temperatures, compared to typical main sequence GOODS-ALMA star-forming galaxies at the same stellar mass and redshift.
They suggest that their star formation rate is somehow sustained in very massive star-forming galaxies, keeping them within the MS even when their gas fractions are low.
They claim that these galaxies are presumably on their way to quiescence. 

To be able to put a constraint on the evolutionary path of these galaxies, we must be able to reconstruct the star formation history (SFH) of galaxies.
In absence of SFH tracers from emission and/or absorption emission lines, SFH recovery must rely on spectral energy distribution (SED) modelling of continuum emission.
Although parametric SFH can be used to derive physical properties of galaxies, they suffer from strong biases preventing a correct reconstruction of the SFH \citep[see for instance,][]{Buat14,Ciesla15,Ciesla17,Carnall19,Lower20,Leja21}. 
Several methods making use of parametric SFHs have been developed to at least put constraints on the last few hundreds Myr of the SFH using, for instance,  Bayesian Information Criterion \citep{Ciesla18} or Approximate Bayesian Computation \citep{Aufort20,Ciesla21}.
However, to go back further in time parametric SFH can no longer be used and more advanced SFH formulation have to be taken into account.
Non-parametric SFHs have been proposed to break free from the degeneracies and biases produced by parametric SFHs \citep[e.g.,][]{Iyer17,Iyer19,Leja19a,Lower20} and have been implemented in several SED modeling code such as \texttt{Prospector} \citep{Leja17,Johnson21} or \texttt{Bagpipes} \citep{Carnall18}.

In this paper we aim at investigating the SFH of GOODS-ALMA galaxies, in particular the ``SB in the MS'' to understand if the extreme properties they exhibit can be explained by the history of their star formation activity.
To reach this goal, we implement in \cigale\ non-parametric SFHs and test their accuracy and sensitivity to recent SFH variations.
We use this approach to reconstruct the last Gyr SFH of GOODS-ALMA galaxies.
This article is structured as follows:
The GOODS-ALMA sample is described in Sect.~\ref{sample}.
The SED modelling procedure, including the addition of non-parametric SFHs and tests on their sensitivity, is detailed in Sect.~\ref{sedmodeling}.
Results from SED modeling and the analysis of the recent SFH of GOODS-ALMA galaxies are provided in Sect.~\ref{results}.
A discussion on these results and the conclusion of the paper are detailed in Sect.~\ref{discussion} and Sect.~\ref{conclusions}, respectively.
Throughout the paper, we used a \cite{Salpeter55} initial mass function.


\section{\label{sample}The GOODS-ALMA sample}
GOODS-ALMA is a blind 1.1\,mm galaxy survey of the deepest part of the Great Observatories Origins Deep Survey South field \citep[GOODS-South;][]{Dickinson03,Giavalisco04}. The survey covers a continuous area of 72.42\,arcmin$^2$ with ALMA Band 6 observations using two array configurations, providing high and low angular resolution datasets at a homogenous average sensitivity (programs 2015.1.00543.S and 2017.1.00755.S; PI: D. Elbaz). The high resolution dataset was presented in \cite{Franco18} (GOODS-ALMA 1.0), while the low resolution dataset and its combination with the high resolution was presented in \cite{GomezGuijarro22a} (GOODS-ALMA 2.0). The combined mosaic reaches an average point-source sensitivity of 68.4$\,\mu$Jy beam$^{-1}$ at an average angular resolution of 0.447\arcsec\,$\times$\,0.418\arcsec. We refer the reader to \cite{Franco18} and \cite{GomezGuijarro22a} for details about the survey observations, data processing, and source catalogue.

In this work, we use the GOODS-ALMA 2.0 source catalogue presented in \cite{GomezGuijarro22a,GomezGuijarro22b} composed of 88 sources. In particular, we focus on the subset of galaxies with a \textit{Herschel} counterpart (69/88), as presented in \cite{GomezGuijarro22b}, but discarding galaxies labelled as optically dark/faint (also known as \textit{HST}-dark) in the GOODS-ALMA 2.0 catalogue. The sample is composed of 65 galaxies with continuous ultraviolet (UV) to millimeter (mm) photometry coverage, which is needed for the purpose of this work. In the following, we adopt the redshifts reported in the GOODS-ALMA 2.0 source catalogue.

UV and near-infrared photometry are taken from the ASTRODEEP-GS43 \citep{Merlin21} catalogue. This catalogue provides updated consistent photometry measurements in 43 optical and IR bands (25 wide and 18 medium filters) ranging from $U$-band to \textit{Spitzer}/IRAC/8\,$\mu$m in the GOODS-South field. Additionally, we employ mid-IR to mm data including \textit{Spitzer}/MIPS/24\,$\mu$m images from GOODS, \textit{Herschel}/PACS/70, 100, 160\,$\mu$m from GOODS-\textit{Herschel} \citep{Elbaz11} and PEP \citep{Lutz11} combined \citep{Magnelli13}, \textit{Herschel}/SPIRE/250, 350, 500\,$\mu$m from HerMES \citep{Oliver12}, and ALMA 1.1\,mm from GOODS-ALMA 2.0 \citep{GomezGuijarro22a}. We refer  the reader to \cite{GomezGuijarro22b} for further details about the photometry at these wavelengths.

\section{\label{sedmodeling}The SED modeling procedure}

\subsection{\label{CIGALE}Inclusion of non-parametric star formation histories in \cigale}

We use the SED modeling code \cigale\footnote{\url{https://cigale.lam.fr/}} \citep{Boquien19} which combines a set of modules modelling the SFH, the emission of the stellar populations, the attenuation from dust, the dust emission, and the AGN contribution from X-ray to radio.
Regarding the SFH, two main approaches are already implemented in \cigale.
The first one is the possibility of using any file containing an SFH, such as the output of a simulation for instance.
The second approaches is to use a parametric model such as exponentially declining/rising or $\tau$-delayed SFHs, adding a recent burst or quenching.
In this study we add a third possibility: a non-parametric set of SFHs.
For this, we use the 2022.1 version of \cigale\ customised for the purpose of this work.

Contrary to the parametric SFHs, the non-parametric SFHs do not assume any analytic function to model the SFH.
This approach assume a given number of time bins in which the SFRs are constant.
We take advantage of the versatility of \cigale\ and implement a new SFH module \textsc{sfhNlevels}.
Based on the results of \cite{Ocvirk06}, \cite{Leja19a} defined 7 redshift bins in lookback time $[0, 30, 100, 330, 1100, 3600, 11700, 13700]$ (in Myr).
The number and time limits of these bins have extensively been tested by \cite{Leja19a} and used in further works \citep[e.g.][]{Leja21,Tacchella20,Lower20,Tacchella21b}. 
Like in the other SFH modules of \cigale, the age of the galaxy is a free parameter.
For each age value, the time interval is divided into 7 bins conserving the logarithmic scaling used in \cite{Leja19a}.
The first bin age is fixed and defined as an input parameter.
The limits of the time bins are thus only dependent on the input values of the age provided to \cigale\ before a run.

To link the SFR of each bin and prevent the computation of unrealistic SFHs, we use a prior which weights against sharp transitions between the SFH bins.
Several studies have tested different priors such as Dirichlet or continuity for instance \citep{Leja19a,Lower20,Tacchella20,Tacchella21b,Suess22}.
We choose to implement in \cigale\ the \textit{bursty continuity} \citep[see for instance][]{Tacchella21b} which, in light of the results of these studies, offers the best compromise in terms of accuracy in retrieving the stellar mass and SFR of galaxies.
In detail, the continuity prior is a Student-t distribution for the prior $x=\log(\mathrm{SFR}_n/\mathrm{SFR}_{n+1})$:

\begin{equation}
    \label{eqn:cont}
    \mathrm{PDF}(x,\nu) = \frac{\Gamma(\frac{\nu+1}{2})}{\sqrt{\nu \pi} \Gamma(\frac{1}{2}\nu)}    \left(1+ \frac{(x/\sigma)^2}{\nu} \right)^{-\frac{\nu+1}{2}},
\end{equation}
where $\Gamma$ is the Gamma function, $\sigma$ a scale factor controlling the width of the distribution, and $\nu$ the degree of freedom controlling the probability in the tails of the distribution.
We choose to adopt $\nu=2$ following \cite{Leja19a} and $\sigma=1$ following \cite{Tacchella21b}. 
Indeed, as shown in \cite{Tacchella21b} this prior will allow to reach a wider range of specific SFR (sSFR) compared to the simple \textit{continuity} prior with $\nu=2$. 
In the \textsc{sfhNlevels} module, the SFR of the first bin is set to 1, then the SFR of the following bins are obtained randomly choosing $x$ in the Students-t distribution.
The obtained SFH is then normalised to 1\,M$_{\odot}$ and used to build the modelled SED.
\cigale\ provides  output  parameters obtained from the best fit models as well as  derived from a Bayesian-like analysis, in other words, computed from the probability distribution function (PDF) of a given parameter.
The SFR in each bin can be obtained from the PDF analysis as well.
The age of formation, $age_{form}$, which is the time at which the galaxy has formed half of its stellar mass is also an output of the module.

\subsection{\label{accuarcy}Accuracy of the non-parametric SFH}

We test the ability of non-parametric SFHs to model the SED of galaxies using a set of 100 SFHs of simulated $z=1$ typical main sequence galaxies obtained with the semi-analytical code \textsc{Galform} \citep{Cole00,Bower06,BensonBower10} already used in \cite{Ciesla15} to test several parametric SFH.
\cigale\ is used to produce SEDs associated to the \textsc{Galform} SFHs assuming different models.
In this test, to focus only on the effect of the SFH model used to perform the SED fitting, we build and fit the SED using the same single stellar population model \citep{BruzualCharlot03}, attenuation law \citep{Calzetti00}, and dust emission \citep{Dale14}.
This test has already been made in \cite{Ciesla15} to quantify the ability of exponential and $\tau$-delayed SFHs to recover the stellar mass and SFR of simulated galaxies.
To be able to compare with their results, we build mock galaxy SEDs associated with the \textsc{Galform} SFHs using the same set of parameters. 
We then run \cigale\ using a $\tau$-delayed plus flexibility SFH as our fiducial parametric SFH.
This model has been proposed in recent studies to decouple the long term SFH from the recent SF activity by allowing an additional flexibility to the $\tau$-delayed SFH \citep{Ciesla17,Schreiber18jekyll,Malek18}.
Then, we run \cigale\ using the non-parametric SFH.

\begin{figure}[ht] 
  	\includegraphics[width=\columnwidth]{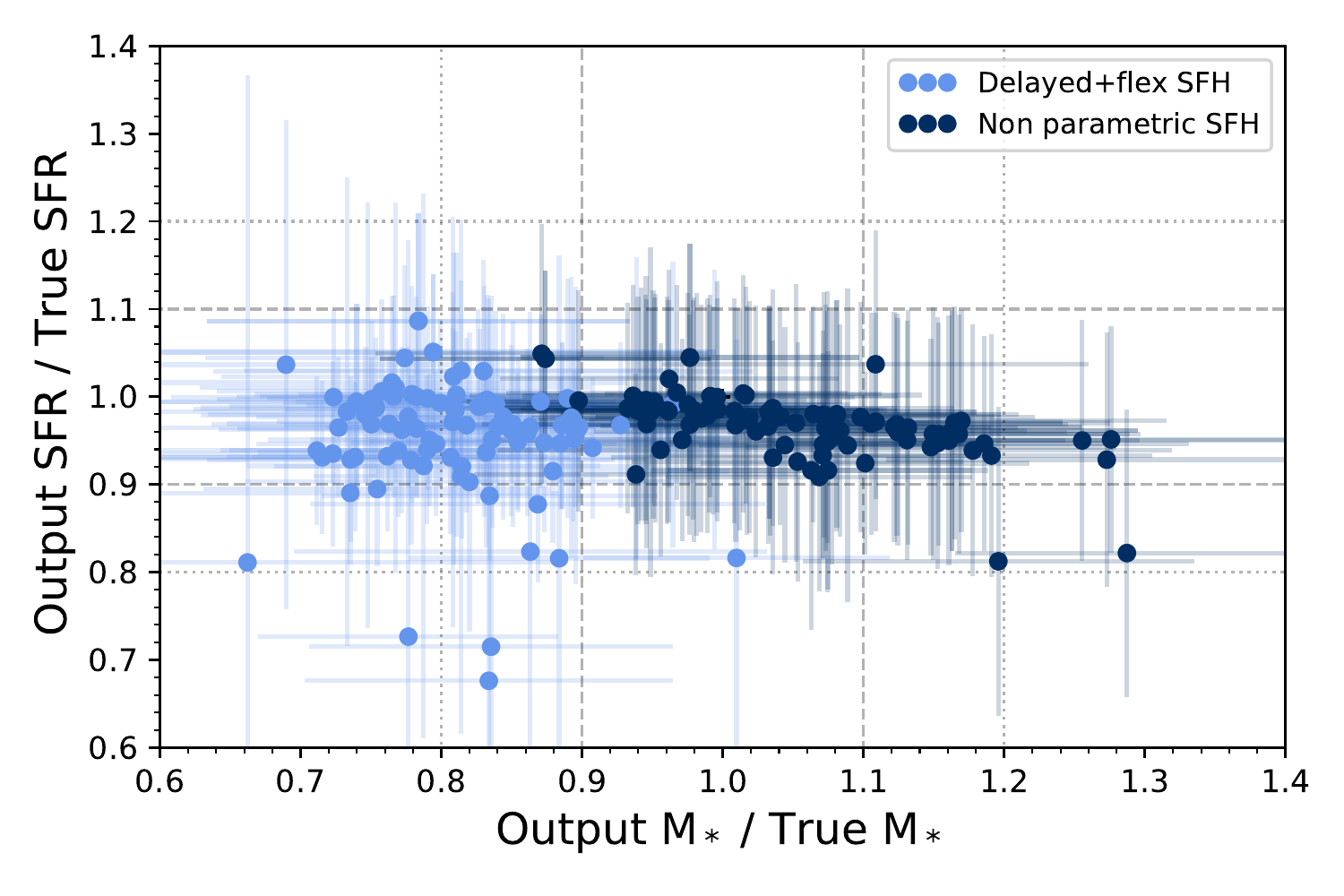}
  	\caption{\label{galform} Comparison between the physical properties (stellar masses and SFRs) obtained by \cigale\ and the true properties of the \textsc{Galform} simulated galaxies. Results are compared using a $\tau$-delayed SFH (light blue points) and non-parametric SFH (dark blue points).}
\end{figure}

The results are shown in Fig.~\ref{galform} for the stellar mass and SFR.
The $\tau$-delayed+flex SFH clearly underestimates the stellar mass of the \textsc{Galform} galaxies with a mean shift of $-18.9 \pm 6.5\%$ while the SFR is well-recovered  with an offset compared to the true value of $-4.0 \pm 6.7\%$.
These results are consistent with \cite{Lower20} and \cite{Iyer17} who also found an underestimation of the stellar mass and a good estimate of the SFR from the $\tau$-delayed + burst SFH.
When using the non-parametric SFH, the offset regarding the stellar mass is more dispersed but more centred around the true value ($3.0 \pm 3.6\%$).
The SFR is well recovered as well with a mean offset of $4.7 \pm 3.3\%$.
These results are consistent with those of \cite{Leja21} who obtained larger stellar masses than SED fitting methods using parametric assumptions for the SFH as well.

We consider now the results of the analysis of \cite{Ciesla15} (see their Table~4) that tested the 1-exponentially declining and 2-exponentially declining SFHs as well as the canonical $\tau$-delayed SFH.
Both the $\tau$-delayed+flex and non-parametric SFHs tested in this work provide good results (less than 5\% error) in terms of SFR compared to the SFHs tested in \cite{Ciesla15}.
However, in terms of stellar mass, the non-parametric SFH yields to the best stellar mass estimates (Fig.~\ref{galform}).
As the \textsc{Galform} SFHs show some strong and rapid variations over the life of the galaxies, changes that are quite drastic, we repeat the test smoothing the \textsc{Galform} SFHs over 500\,Myr and 1\,Gyr and found consistent results.
The results of these tests show that the non-parametric SFHs provide a more accurate estimate of the stellar mass of galaxies than parametric models which tend to provide underestimated results.

\subsection{\label{sensitivity}Sensitivity of the non-parametric SFH}

In this work we aim at identifying any recent variation of SF activity that GOODS-ALMA galaxies may have underwent in the last few hundreds of Myr.
Therefore, we need to assess the ability of the non-parametric SFHs to recover such variations.
When introducing the non-parametric SFHs, \cite{Leja19a} tested them on several simple cases including a burst and a sudden quench.
However, in their test the sudden quench occurred 1\,Gyr ago and they found that the continuity prior provides a good estimate of the SFR.
We try to refine the test to see the limits of the SFH sensitivity.
Since we adopt the bursty continuity prior \citep{Tacchella21a} we test how well we can recover this burstiness.

\begin{figure}[ht] 
  	\includegraphics[width=\columnwidth]{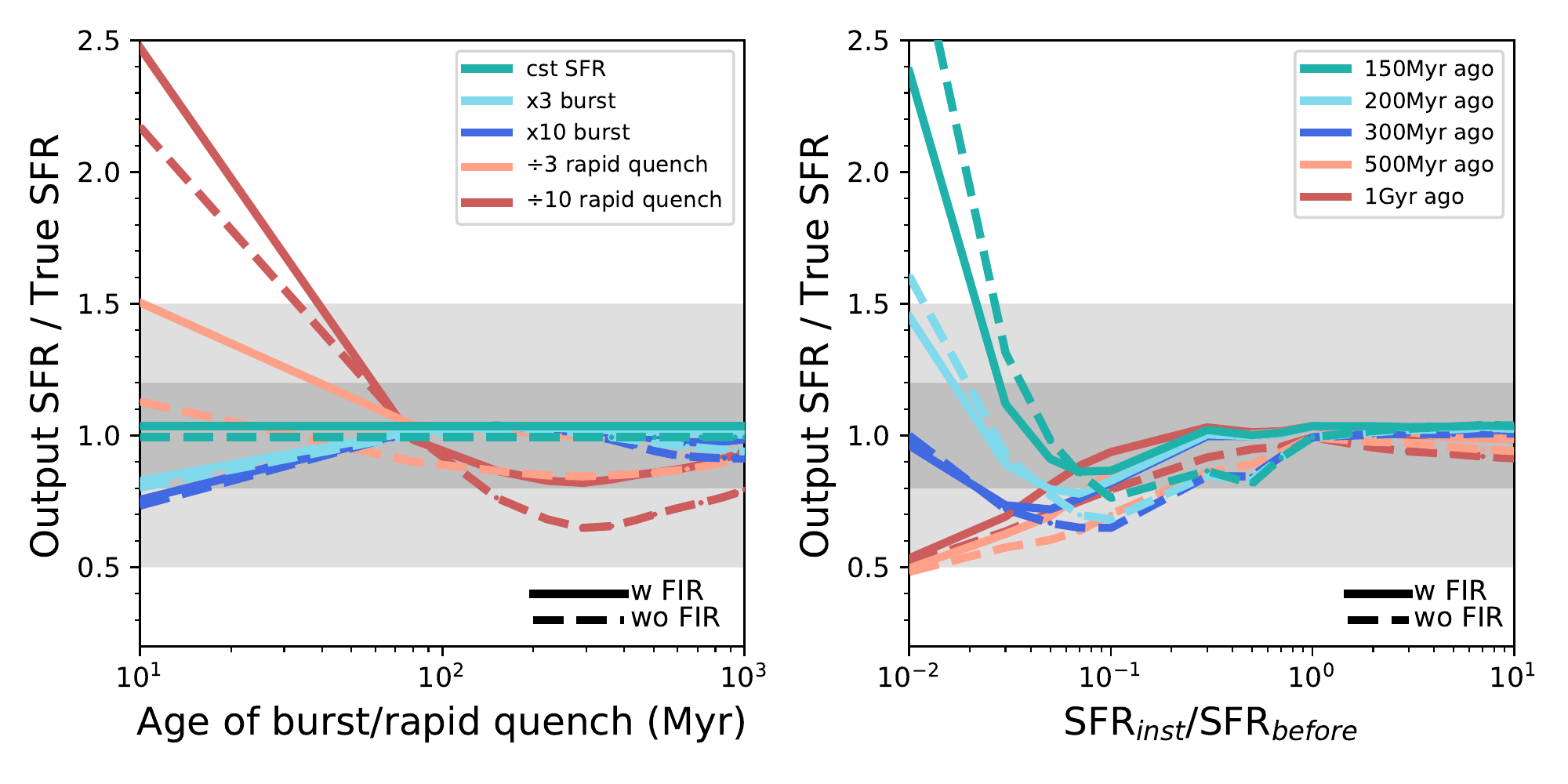}
  	\caption{\label{sensitivity} Sensitivity of the non-parametric SFH in recovering rapid and recent variation of SFH. On both panels, the y-axis indicates the ratio between the estimated SFR obtained from the non-parametric SFH and the true one. The variation of this ratio is shown as a function of the age of the rapid variation (left panel) and its strength, that is the ratio between the instantaneous SFR and the SFR before the burst or quenching (right panel). In both panels, the solid line represents the results obtained when the full observed UV-mm SED is fitted and the dashed line when no FIR data is used for the fit.}
\end{figure}

To perform our test, we simulate a set of mock SEDs with \cigale. 
We assume a constant SFH with a final instantaneous burst or quenching.
The time when this instantaneous variation occurs varies from 10\,Myr to 1\,Gyr.
Its strength, that is the ratio between the actual SFR and the SFR just before the variation, is set between 0.01 to 10, to probe both strong quenching and strong starburst.
The mocks SEDs are then integrated into the GOODS-ALMA set of filters and fluxes are randomly perturbed using a Gaussian distribution with a standard deviation of 10\% of the flux density in each filter.
We then use \cigale\ to fit this mock catalogue using non-parametric SFH and show the results of this test in Fig.~\ref{sensitivity}.
If the SFH is constant or just underwent a rapid burst, even a factor 10 starburst, the SFR is well recovered, whenever this burst occurred.
For starburst events, we note though that the recovered SFR starts to be underestimated between a few to 20\% if this variation happened less that 30\,Myr ago.
Regarding rapid quenching events, the SFR is slightly underestimated if the quenching happened between 100\,Myr and 1\,Gyr. 
Below 100\,Myr, the SFR starts to be overestimated by 50\% if the quenching happened between 40 and 100\,Myr ago, and by a factor up to 2.5 for more recent quenching events.
We note that for these variations, the true SFR is set to be close to 0 and a factor 2.5 overestimate keeps this value very low and therefore do not bias toward active SFR.
We perform this test with and without using the FIR filters.
We find no significant difference in the results if the IR filters are not used.

\subsection{\label{gradients}Definition of the SFR gradient}

\begin{figure}[ht] 
  	\includegraphics[width=\columnwidth]{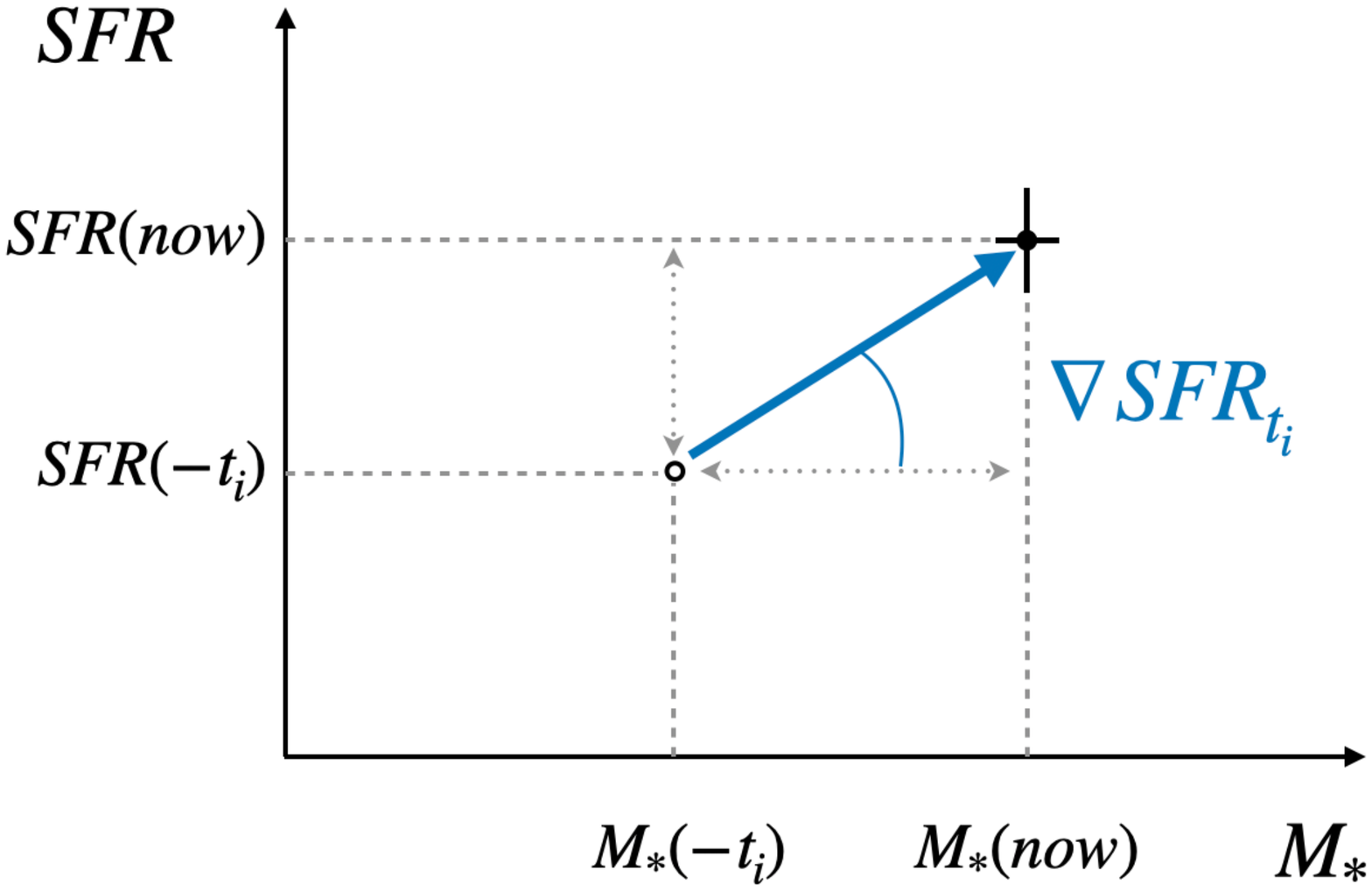}
  	\caption{\label{defgrad} Diagram explaining the definition of the SFR gradient. Using the output SFH from \cigale, the SFR and M$_*$ obtained $t_i$\,Myr ago are used to place the galaxy at the position where it was $t_i$\,Myr ago on the SFR-M$_*$ plane. The SFR gradient is the angle between the line linking the position of the galaxy $t_i$\,Myr ago and its position now and the line drawn from a constant SFR.  }
\end{figure}

For the purpose of this work, a new parameter, the SFR gradient, $\nabla$SFR, is also computed.
This parameter provides the angle showing the direction that a galaxy has followed in the usual representation of the MS, that is the SFR vs stellar mass plane, during a given time (this time is an input free parameter) associated to the modelled SFH.
In other words, the SFR gradient is the angle between the line linking the position of the galaxy $t_i$\,Myr ago and its position now and the line drawn from a constant SFR (Fig.~\ref{defgrad}).
A value close to 0$\degre$ means that the recent SFH is overall constant with no sharp variation.
However a large positive/negative value would mean that the galaxy undergoes a starburst/quenching phase.
The SFR gradient allows us to analyse quantitatively the recent SFH of galaxies and see galaxies' movements relative to the MS.

\section{\label{results}Results from SED modeling}

We run \cigale\ on the GOODS-ALMA sample using the \textsc{sfhNlevels} SFH module, stellar populations models of \cite{BruzualCharlot03}, a modified \cite{Calzetti00} starburst attenuation law module, and the \cite{Dale14} IR templates.
The parameters used for the two runs are provided in Table~\ref{inputparam}.
The code uses energy balance to fit the entire SED from UV to submm as we rely on the results of \cite{Seille22} who tested it intensively on the highly perturbed and star-forming Antennae system.
They showed that the results of the UV-submm SED modelling of the entire system, using energy balance, is consistent with the sum of the results of the UV-submm SED modelling of individual regions although they exhibit highly different level of attenuation.

\begin{table}
   \centering
   \caption{\cigale\ input parameters used to fit the GOODS-ALMA sample. This set of parameters resulted in 378,000 SED models.}
   \begin{tabular}{l c l}
   \hline\hline
   \textbf{Parameter} & \textbf{Value} & \\
   \hline
   \multicolumn{3}{c}{\textbf{Non-parametric SFH --} \sc{sfhNlevels}}\\[1mm]  
   $age$ (Gyr) & $\big[$1.2;\ 8.7$\big]$     & 10 values linearly sampled   \\
   N$_{SFH}$ & 10,000 & \# of SFH per $age$ value \\[1mm]
   First bin age & 30\,Myr & \\[1mm]
   \multicolumn{3}{c}{\textbf{Dust attenuation--} \sc{dustatt\_modified\_starburst}}\\[1mm]  
   E(B-V)s lines      &  $\big[$0;\ 1.5$\big]$     & 14 values linearly sampled  \\
   E(B-V)s factor      &  1     &   \\
   \multicolumn{3}{c}{\textbf{Dust emission --} \cite{Dale14}}\\[1mm]  
   $\alpha$      &  1, 1.5, 2, 2.5, 3     &   \\
   \hline
   \label{inputparam}
   \end{tabular}
\end{table}

\cigale\ provides as results the output values of the parameters computed from the Bayesian-like analysis. 
The output value of a parameter is the mean of its PDF, assuming that it is Gaussian, and the error is the standard deviation of the PDF.
Parameters linked to the SFH are known to be difficult to constrain \citep[see for instance,][]{Buat15,Ciesla17,Carnall19}.
Therefore, to go a step further and obtain more robust results, we perform an additional step in our estimate of the physical parameters of the GOODS-ALMA galaxies.
For each galaxy, we add random noise assuming a Gaussian distribution with $\sigma=0.1 \times S_{\nu}$ to the flux of each band, and repeat the operation 10 times.
Therefore, each galaxy will be fitted 10 times with \cigale.
The final value of a parameter is the mean value of the 10 Bayesian output values, and the error is the dispersion of these 10 Bayesian output values.

\subsection{\label{mocksanalysis}Mock analysis}

\begin{figure*}[ht] 
  	\includegraphics[width=\textwidth]{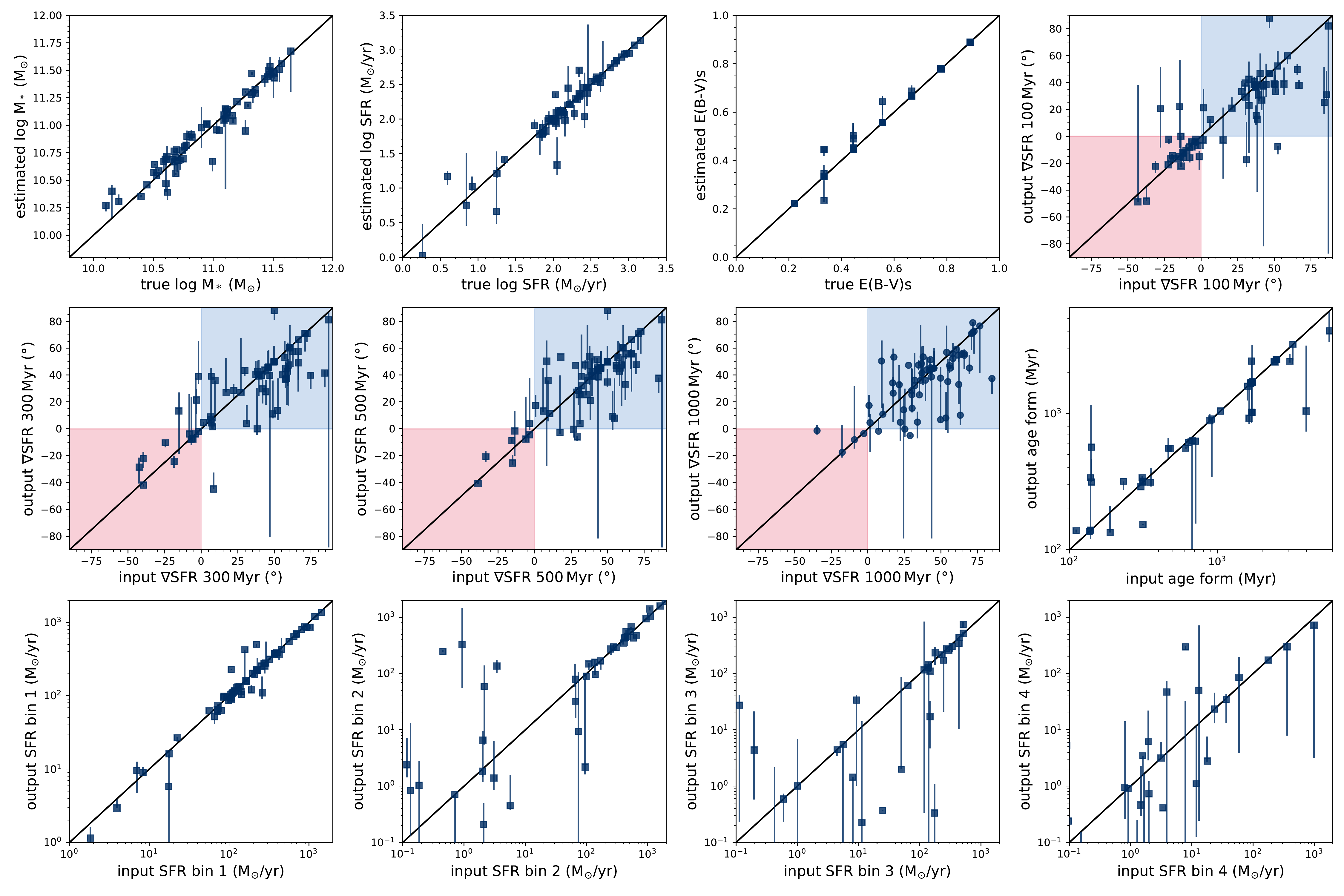}
  	\caption{\label{mocks_ir} Results of the mock analysis built from the GOODS-ALMA galaxies for the key parameters used in this work. The black solid line is the one-to-one relationship. }
\end{figure*}

As a standard procedure, we analyse the ability of the code to constrain the key parameters necessary to this study.
The mock analysis procedure has been extensively described in previous works \citep[e.g.][]{Buat19,Boquien19,Ciesla21}, its 
Although its principle is the same than the tests performed in Sect.~\ref{sensitivity} and Sect.~\ref{accuarcy}, the difference here is that the mock SEDs are created using the SED fit of the GOODS-ALMA galaxies of our sample.
The best models of this first run are used as mock galaxies representative of the GOODS-ALMA population for which all parameters are known.
Thus, the input parameters are representative of our galaxy population.
A second run of \cigale\ is used to see how well we constrain these parameters.
The results of the mock analysis are shown in Fig.~\ref{mocks_ir} for the key parameters used in this study.
The output values are computed using the new method just described in the previous section.
A one-to-one relationship between the true values and the estimated ones implies a perfect constraint of the parameter.

As expected given the wavelength range covered by the GOODS-ALMA set of filters, the stellar mass, SFR, and dust attenuation are well recovered.
Regarding the parameters linked to the SFH, the formation age of the galaxies is relatively well constrained.
Usually, the age of the galaxies is a parameter difficult to constrain and can be found fixed in several papers in the literature \citep[see for instance,][]{Ciesla21}.
The SFR in the first SFH bin is well constrained while it is more difficult for the SFR in the older bins. 
In the second and third bins of SFH, only the high SFR values are well recovered. 
The SFR of bin 4 seems to be constrained as well, although the relation is more dispersed.
Regarding the SFR gradient, the relations are dispersed but do not show strong bias.
Positive gradients are recovered positive and the same goes for negative values.
A few galaxies show some discrepancies, however, for the purpose of this work being able to differentiate between increasing and decreasing SF activities is sufficient.
In conclusion, the constraints on the SFR, M$_*$, and SFR gradients allow us to conduct our analysis.

\subsection{Physical parameters}
\begin{figure}[ht] 
  	\includegraphics[width=\columnwidth]{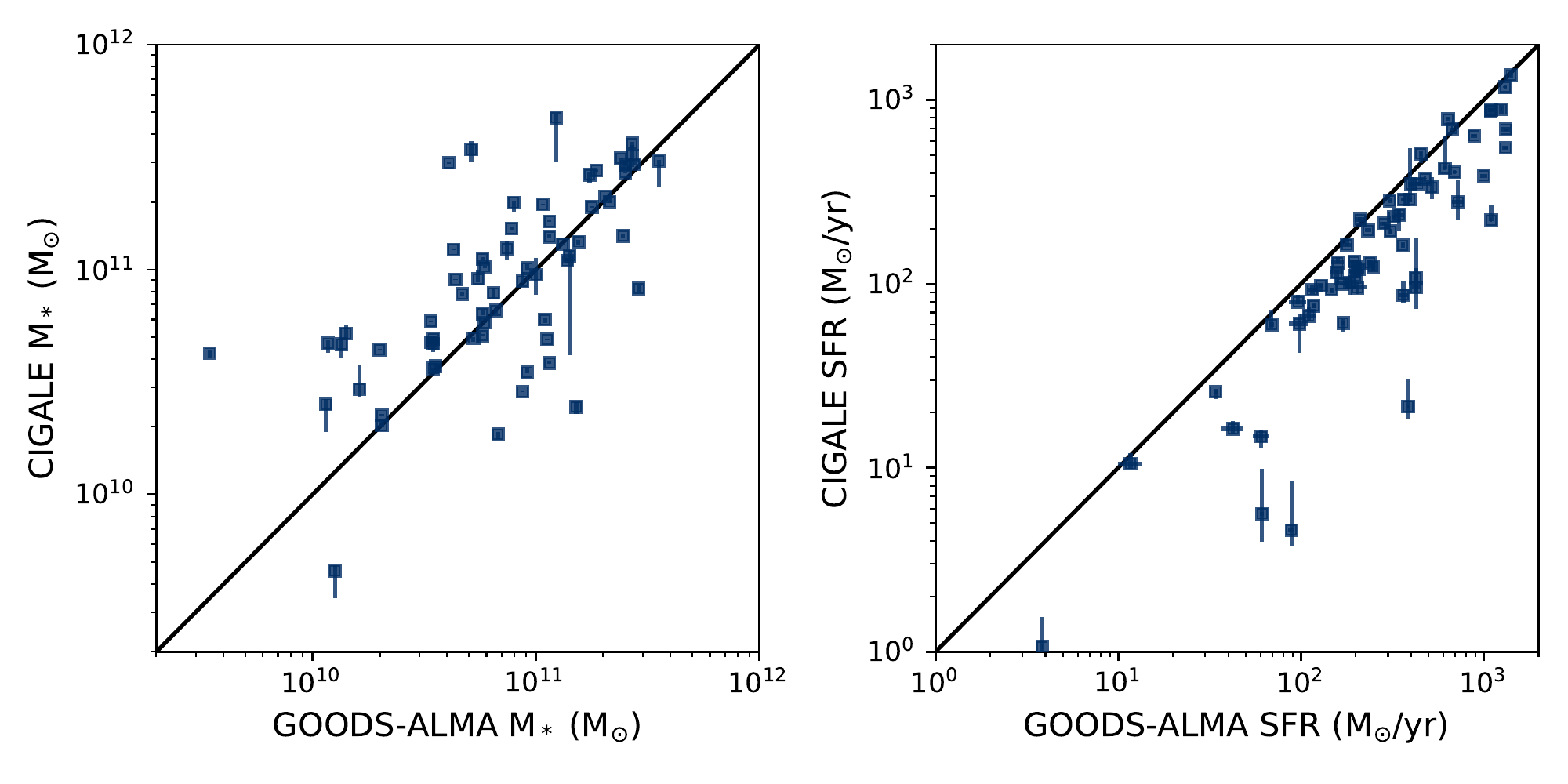}
  	\caption{\label{compgacig} Comparison between the stellar mass and SFR values of the GOODS-ALMA sample provided in \cite{GomezGuijarro22b} and the ones obtained with \cigale\ using a non-parametric SFH. }
\end{figure}

All redshifts included, from 0.12 to 4.73, the mean stellar M$_*$ estimated with \cigale\ is $1.3\times 10^{11}$\,M$_{\odot}$, with values ranging from $1.1\times 10^{9}$ to $4.7\times 10^{11}$\,M$_{\odot}$.
We compare our stellar masses and SFR estimates to the values used in \cite{GomezGuijarro22b} in Fig.~\ref{compgacig}.
There is a significant scatter in the estimate of the stellar mass.
On average, we find larger stellar masses with a median ratio of 1.26, which is expected, and a standard deviation of 2.14.
Indeed, \cite{Leja21} showed that using non-parametric SFH results in larger stellar mass estimates.
This is due to the fact that parametric SFH are rigid and in order to provide good fit and estimates of the SFR, the resulting SFH are younger \cite[see also,][]{Ciesla17,Carnall19}.
Our results on GOODS-ALMA galaxies confirm what we obtained from simulations in Sect.~\ref{CIGALE} and the results of \cite{Leja21}.
Regarding the SFR, it is on average lower than the estimates from \cite{GomezGuijarro22b} with a median ratio of 0.40 and a dispersion of 0.35.
This is due to the fact that \cigale\ takes into account the IR emission due to dust heated by older stellar populations, resulting in lower SFR compared to UV+IR inferred values, a conclusion also reached by \cite{Leja21}.

\subsection{\label{sfhga}The recent SFH of GOODS-ALMA galaxies}

We now look at the recent SFH obtained for the GOODS-ALMA galaxies.
We emphasise that the SFRs and stellar masses used from now are those obtained by the \cigale\ SED fitting and not the values used by \cite{GomezGuijarro22b}.
However, we keep the ``SB in the MS" sample defined by \cite{GomezGuijarro22b} to discuss the SFH properties of these particular sources.

\begin{figure*}[ht] 
    \begin{center}
  	\includegraphics[width=5in]{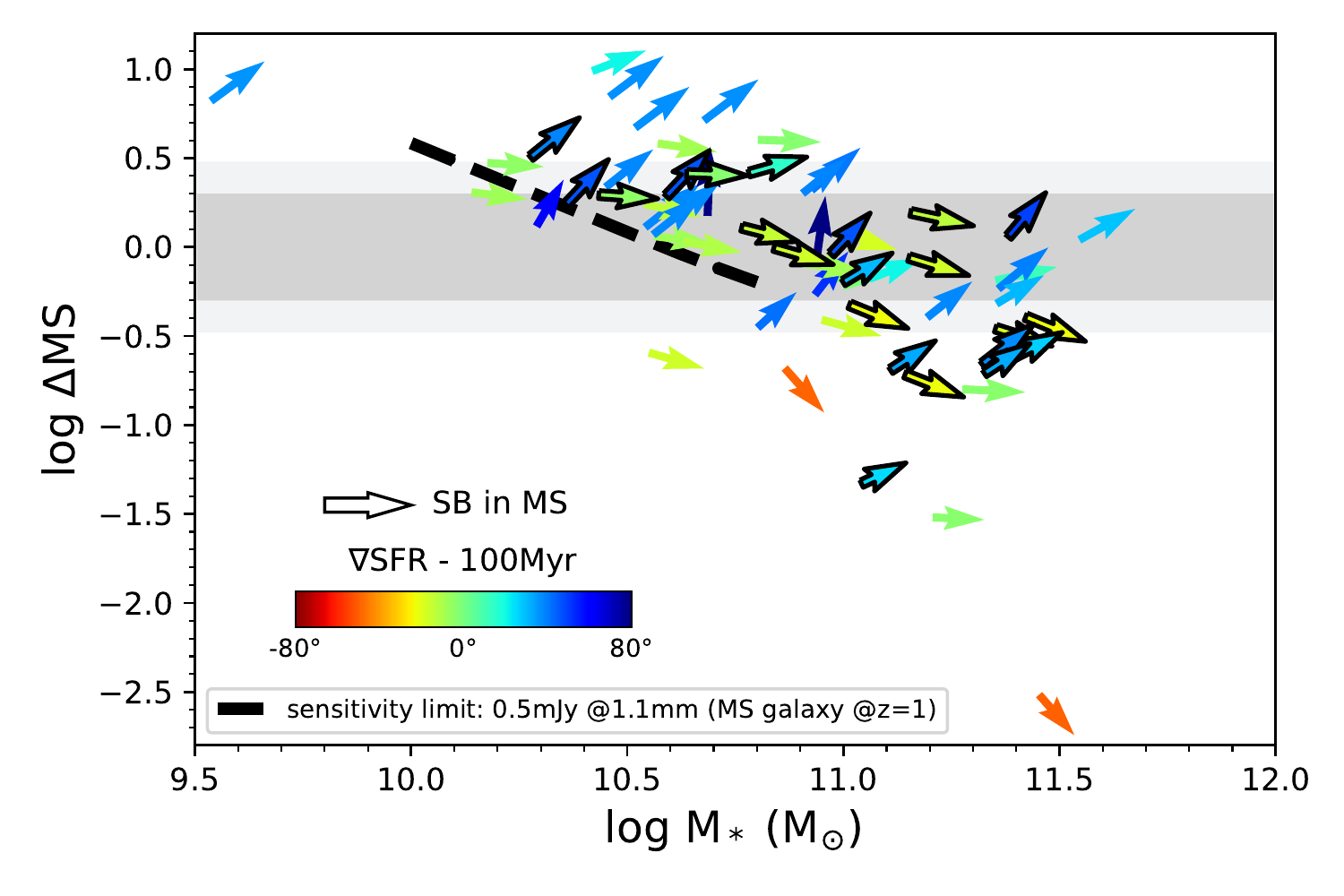}\\
  	\includegraphics[width=\columnwidth]{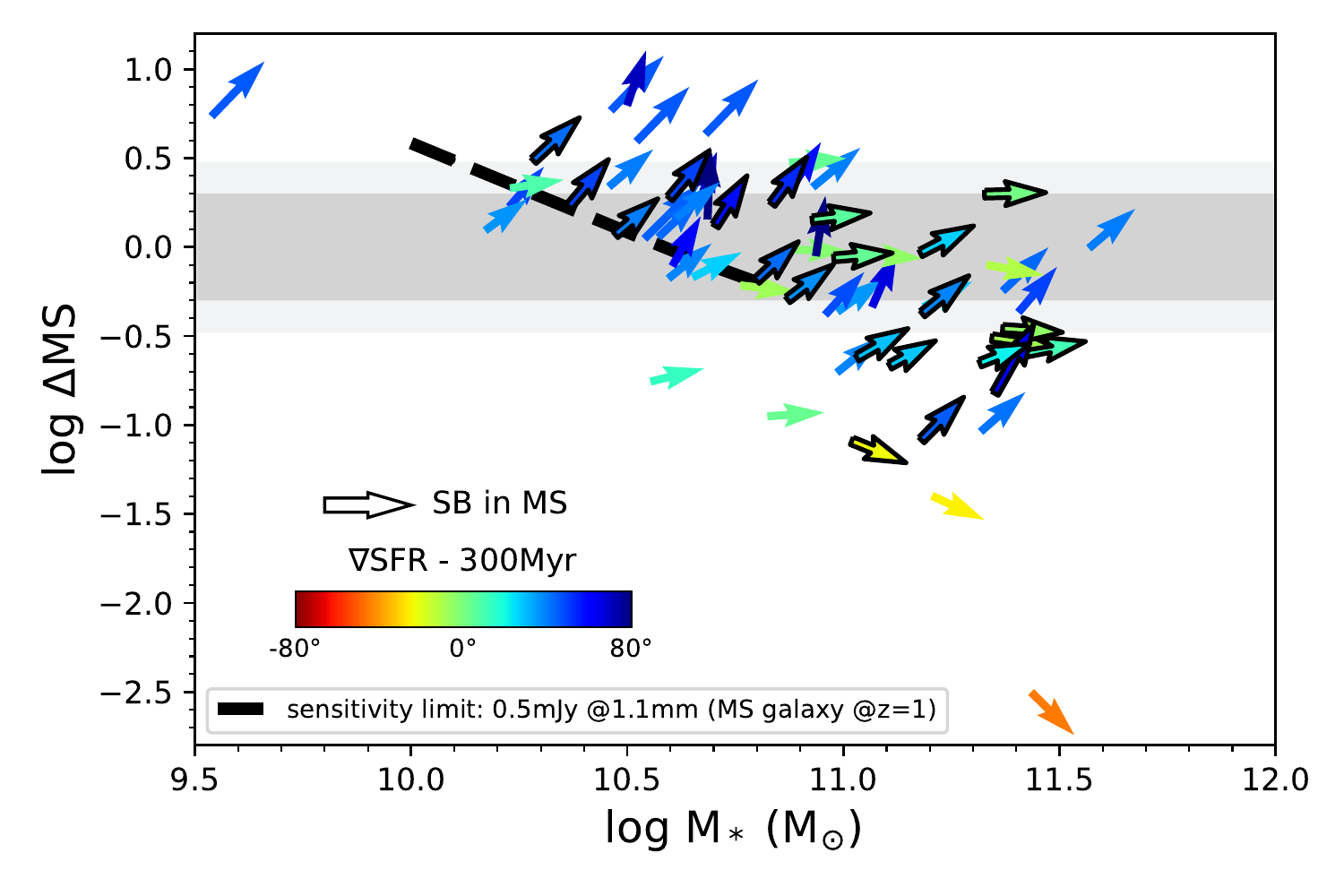}
  	\includegraphics[width=\columnwidth]{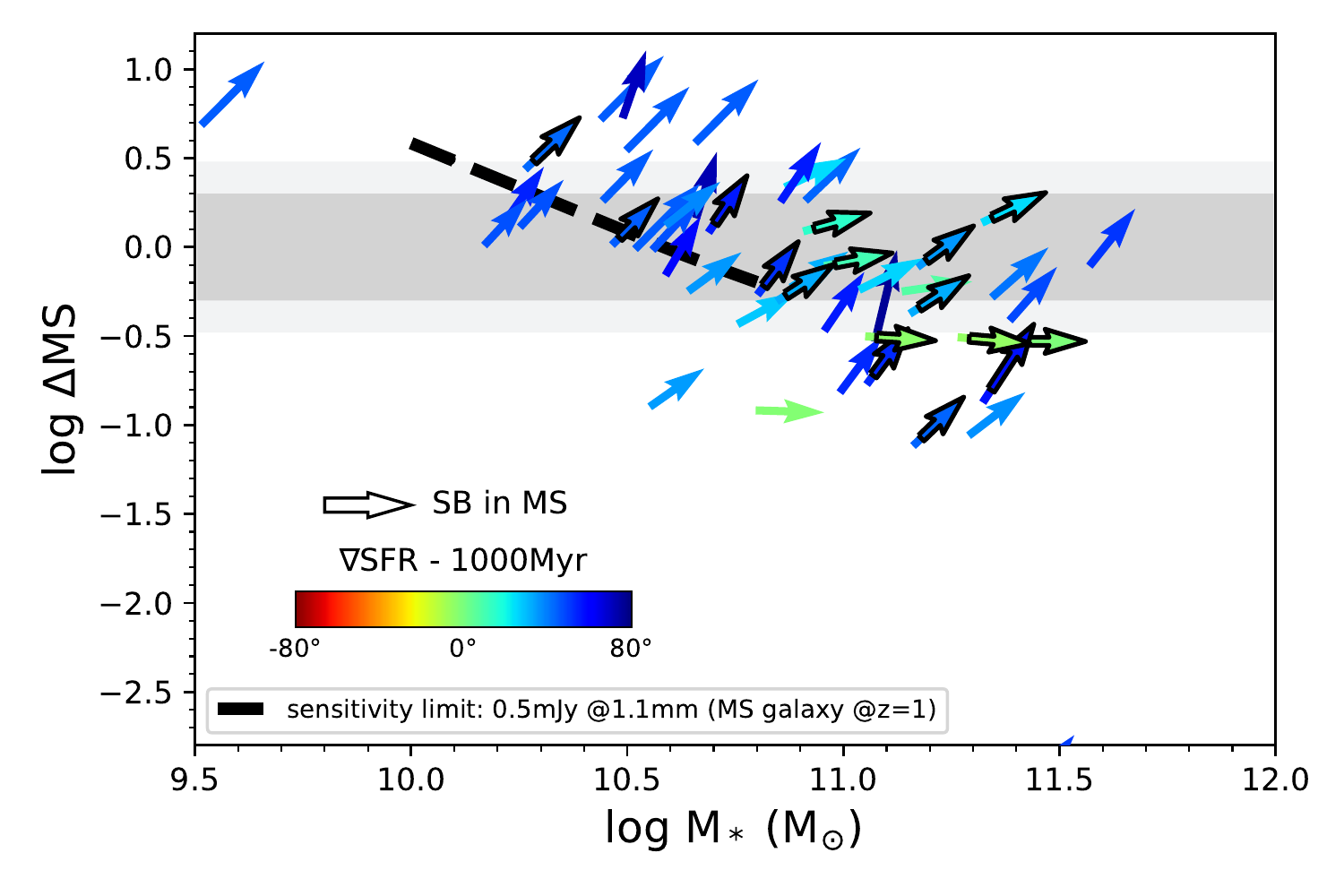}
  	\end{center}
  	\caption{\label{msgradient} GOODS-ALMA galaxies, including ``Sb in the MS'' (with black coloured edges), placed on a $\Delta$MS-M$_*$ plane. The assumed MS is from \cite{Schreiber15}. The arrows orientation and colour indicate the SFR gradient of each galaxies over a given time: 100\,Myrs (top panel), 300\,Myr (bottom left panel), and 1\,Gyr (bottom right panel).}
\end{figure*}

\begin{figure*}[ht] 
    \begin{center}
  	\includegraphics[width=0.33\textwidth]{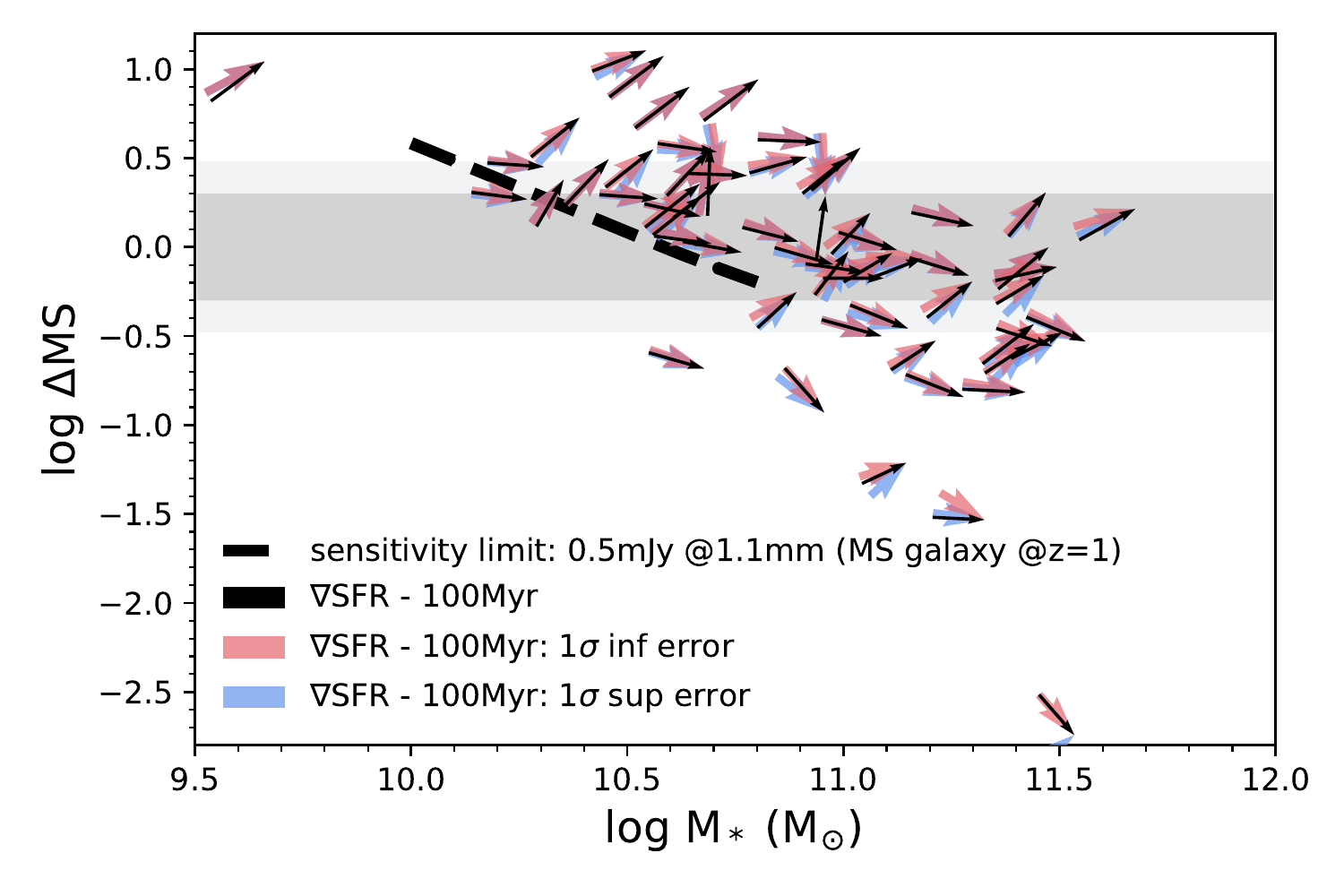}
  	\includegraphics[width=0.33\textwidth]{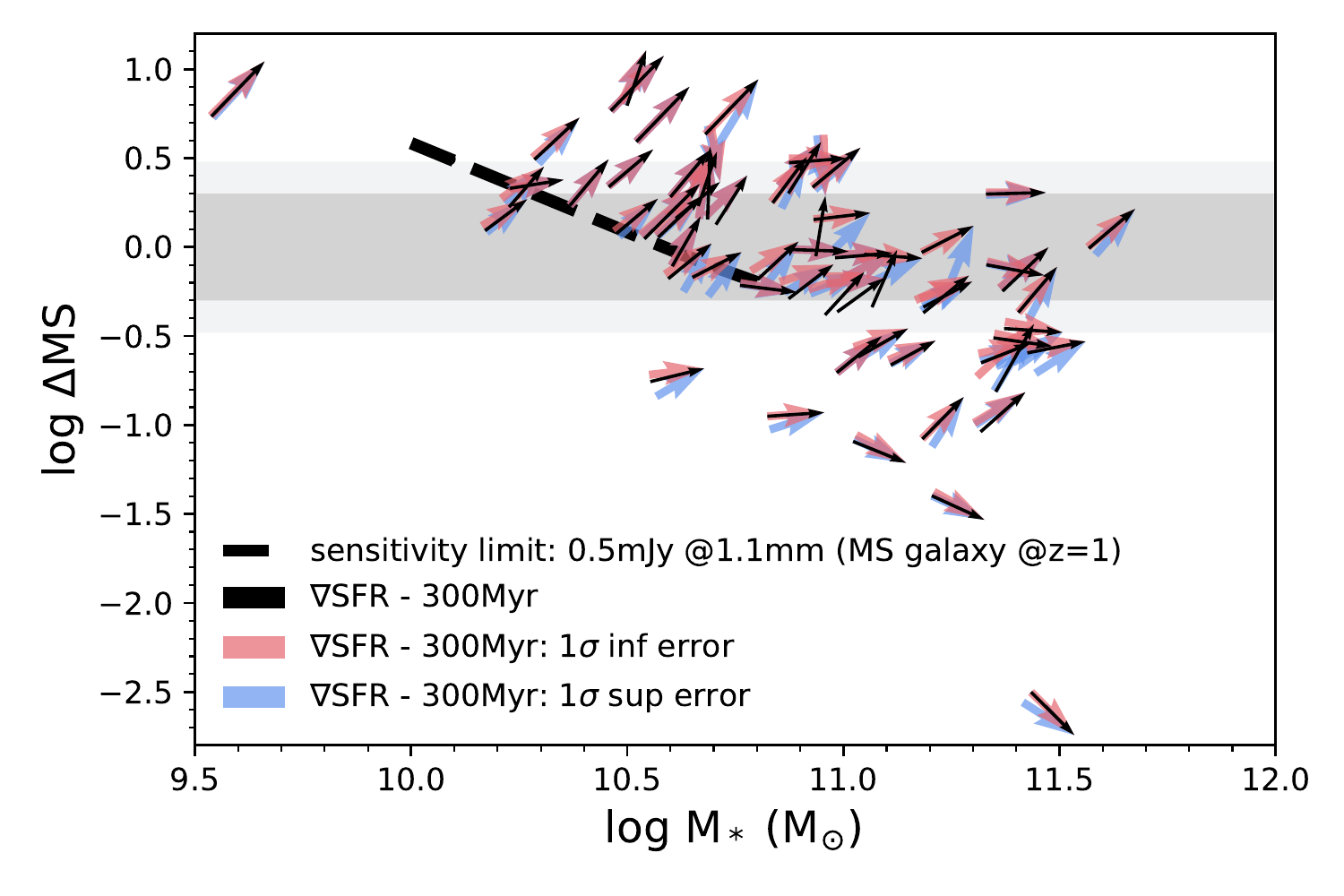}
  	\includegraphics[width=0.33\textwidth]{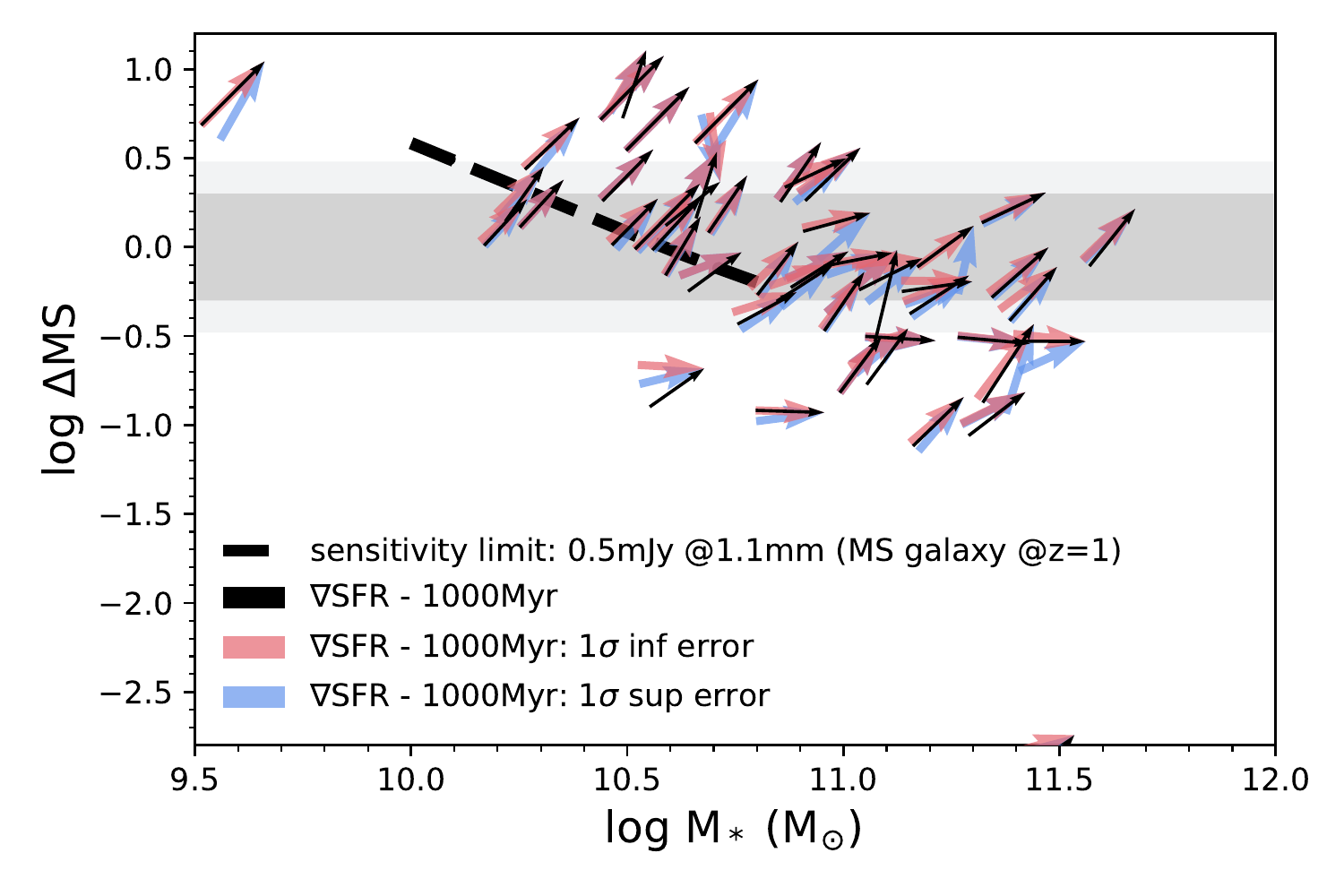}
  	\end{center}
  	\caption{\label{msgradient_err} GOODS-ALMA galaxies, including ``Sb in the MS'', placed on a $\Delta$MS-M$_*$ plane. The assumed MS is from \cite{Schreiber15}. The thin black arrows orientation indicate the SFR gradient of each galaxies over a given time: 100\,Myrs (left panel), 300\,Myr (middle panel), and 1\,Gyr (right panel). The red and blue arrows represent the upper (blue) and lower (red) 1$\sigma$ error on the gradient estimates.}
\end{figure*}

In Fig.~\ref{msgradient}, we place these sources relatively to the \cite{Schreiber15} MS as a function of their stellar mass.
Galaxies are colour-coded according to their SFR gradient obtained over 100, 300, and 1000\,Myr.
The arrows are a visual indication of these gradients, those with a black edge indicates galaxies classified as SB in the MS by \cite{GomezGuijarro22b}.
These arrows allow to give a hint of the movement of galaxies on the SFR-M$_*$ plane.
On top panel, gradients are measured over 100\,Myr.
The GOODS-ALMA galaxies exhibit a wide range of gradients, mostly larger than -25$\degre$.
This timescale reveals the stochastic variations of the recent SFH.
We can distinguish mainly two groups separated in mass.
Galaxies of the first group, with masses lower than 5$\times$10$^{10}$\,M$_{\odot}$, are mostly located in the upper part of the MS, in the light grey region.
No galaxies are found in the centre of the MS scatter or in the lower part.
This is due to the detection limit implied by the GOODS-ALMA blind survey.
Four of these galaxies are located in the SB region, at a factor of 5 above the MS.
This ``low" mass group of galaxies have either a positive gradient, indicative of an undergoing enhanced star-formation activity.
However, a few of them exhibit a flatter one that could indicate a slow decline or constant star-formation in the last 100\,Myr.
Galaxies defined as ``SB in the MS" display both strong and flatter gradients.
Using the SFR estimated by \cigale\ with the non-parametric SFH, they are all located within the MS except one of them.
At higher masses, above 5$\times$10$^{10}$\,M$_{\odot}$, the GOODS-ALMA galaxies lie within the MS, its lower part, or below.
A wider spread of gradient is observed among these sources showing a large range of star-formation histories.
Galaxies classified as ``SB in the MS" share this wide distribution of gradients with the normal GOODS-ALMA galaxies.
Errors on the discussed gradients are provided in Fig.~\ref{msgradient_err} where we show the 1$\sigma$ superior and inferior errors on the gradients.
They are the errors seen in Fig.~\ref{mocks_ir}.
The results of Fig.~\ref{msgradient} are not affected by the errors on the estimates of the gradients parameters.

\begin{figure*}[ht] 
  	\includegraphics[width=\textwidth]{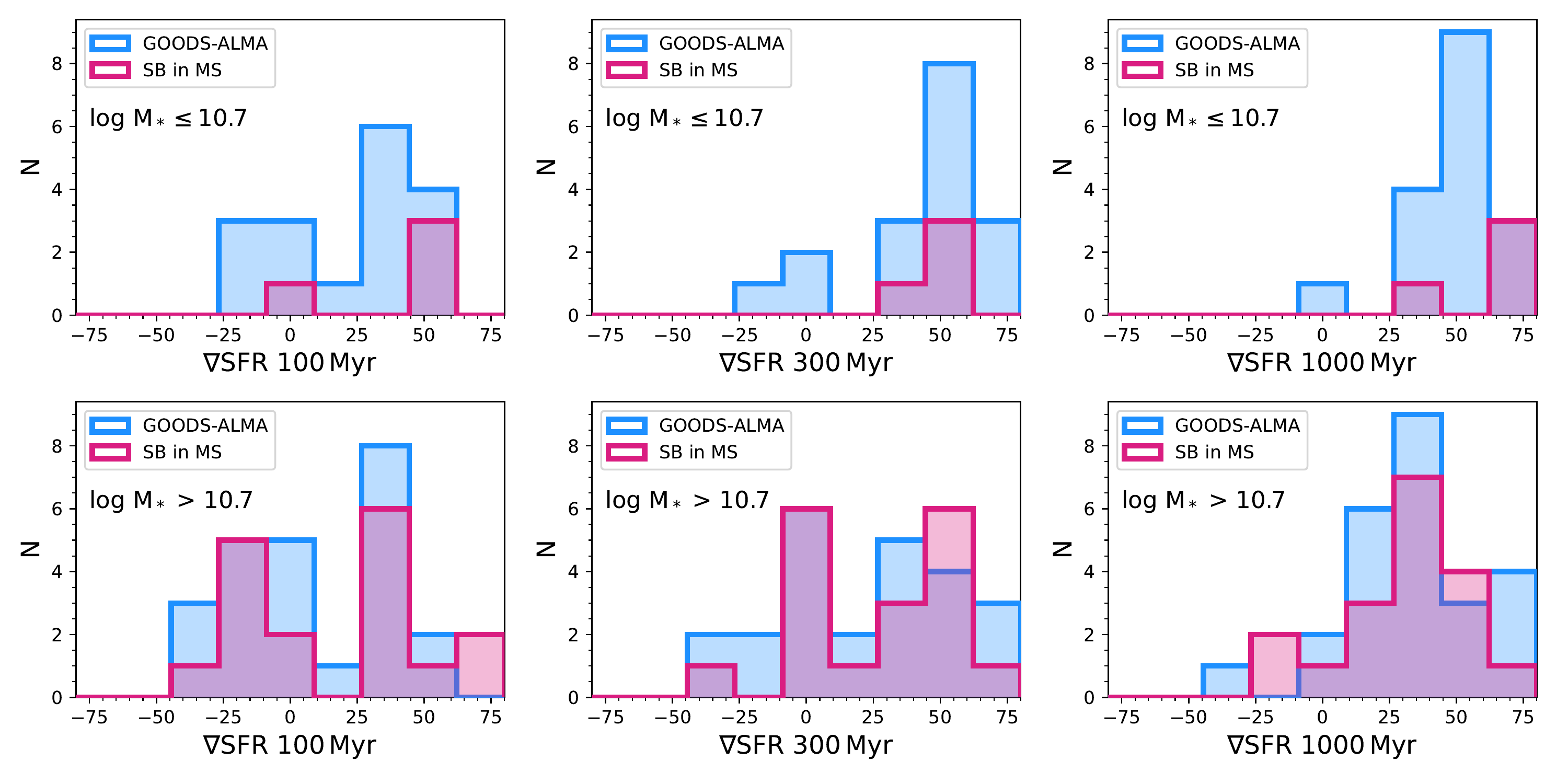}
  	\caption{\label{histsgradient} Histograms of the SFR gradients measured at 100, 300, and 1000\,Myr for the GOODS-ALMA galaxies classified as starbursts in the MS (red) and the other galaxies of the GOODS-ALMA sample (blue). Top row of panels show the distribution for galaxies with $\log M_* \leq 10.7$ while the bottom row panels are galaxies with $\log M_* > 10.7$. }
\end{figure*}

In the GOODS-ALMA sample, only two galaxies have a very low gradient, indicative of a rapid decline of star-formation, that would be expected if a galaxy became rapidly passive.
All the GOODS-ALMA galaxies have a gradient higher than -50$\degre$ as shown if the left panels of Fig.~\ref{histsgradient}.
Their last 100\,Myr SFH do not show any evidence of a rapid decline of the star-formation activity that could have followed a strong starburst phase where the galaxies could have lied above the MS. 
Over a longer timescale, 300\,Myr, arrows indicate mostly an increasing SF activity, with a few galaxies with flatter arrow and decreasing activity.
When the gradients are computed over 1\,Gyr, almost all of the galaxies have an increasing gradient meaning that overall, during the last 1\,Gyr, the SF activity of the GOODS-ALMA galaxies has increased.
Over these two timescales (300\,Myr and 1\,Gyr), no strong decline of the SFH of the GOODS-ALMA galaxies is observed through their SFR gradient.

In Fig.~\ref{msgradient} the compact galaxies, called ``SB in the MS", do not show any specific trend compared to the other GOODS-ALMA galaxies.
To verify this, we show in Fig.~\ref{histsgradient} the 100, 300, and 1000\,Myr gradient distributions of both ``SB in the MS" and the rest of the GOODS-ALMA galaxies, and considering the low (top row) and high (bottom row) mass groups.
For each timescale, there is no significant difference between the gradient distributions of GOODS-ALMA galaxies and SB in the MS, as confirmed by Kolmogorov–Smirnov (KS) tests, whether we consider the low or high mass group.
We conclude that the SB in the MS do not exhibit any particular SFH that could explain their low depletion time compared to the whole GOODS-ALMA sample.

\section{\label{discussion}Discussion}

The motivation behind this work is to try to disentangle between three evolutionary scenarios that \cite{GomezGuijarro22b} proposed to explain the physical properties of the compact galaxies ``SB in the MS''.
The first scenario was introduced in \cite{Tacchella16} and is characterised by compaction events triggered by a strong gas inflow.
If the gas stream is counter-rotating, or if minor mergers are involved, there is a loss of angular momentum.
The inflow rate is more efficient than the SFR, the gas conveys to the centre of the galaxy, and a compact massive core of gas grow yielding a sustained star formation rate.
The compaction phase is characterised by a short depletion time and a high gas fraction. 
As a consequence, the galaxy moves up to the highest part of the MS.
This enhanced star formation activity yields to central gas depletion, and thus inside-out quenching.
The galaxy moves to the lower part of the MS.
Depending on the mass of the dark matter halo in which the galaxy resides, another compaction event can occur, in low mass halos, or, if massive enough, the hot dark matter halo prevents further gas inflow, leading to gas depletion and full quenching.
However, \cite{GomezGuijarro22b} ruled out this scenario to explain the properties of ``starbursts in the MS'' as they do not find galaxies characterised by ongoing compact star formation, short depletion times, high gas fractions, and located within the scatter of the MS, as predicted in this first scenario in the upper bound of the MS.
The favoured scenario is a mechanism that reduces the angular momentum. 
Two possibilities have been considered, one in which the galaxy rises much above the MS (scenario 2) and one where it only mildly rises, remaining within the MS (scenario 3). 
The authors note that existing data discussed in this paper did not allow them to clearly disentangle both scenarios, while scenario 3 could in principle explain why such compact star-formation is systematically found among high-mass MS galaxies.
This evolutionary path, consistent with a slow downfall \citep{Schreiber15,GomezGuijarro19,Franco20}, would imply that the galaxy is still experiencing the last phases of regulation implied by the MS while it is on its way to quiescence.

In the previous section, we showed that GOODS-ALMA galaxies, and in particular the ``SB in the MS'' do not exhibit very low SFR gradients over the last 100\,Myr. 
Scenario 2 of \cite{GomezGuijarro22b} proposes that the low gas fraction of ``SB in the MS" galaxies can be explained by the fact that the galaxies exhausted their gas reservoir during a strong starburst phase putting them above the MS.
If indeed the ``SB in the MS" would have been spotted in the transition between the SB region and the passive one, due to gas exhaustion, one would have expected a very low SFR gradient for these sources.
However, we observe positive or weak negative values for the ``SB in the MS'' galaxies in the last 100\,Myr, which is not compatible with scenario 2.
Even considering a larger timescale, that is 300\,Myr for instance, no sign of rapid quenching is seen among this population neither.
This is also confirmed when looking at the last Gyr.
Therefore, our study rules out scenario 2 of \cite{GomezGuijarro22b}.
We note that \cite{Fensch17} showed that gas-rich mergers do not generate strong starbursts, i.e. with large sSFR much above the MS, due to their strong internal turbulence provoked by their large gas masses. 
In this perspective, one may understand why mergers among gas-rich galaxies within the MS, could reduce the angular momentum of the gas without bringing the galaxies much above the MS as in scenario 2. 
And indeed, it has been claimed that galaxies above the MS at z$\sim$2 systematically exhibit enhanced gas fractions associated with mild increase of SFE although it is still debated \citep[e.g.,][]{Tacconi18}.

Furthermore, we show that ``SB in the MS'' do not exhibit different SFHs than the other galaxies of the GOODS-ALMA sample. 
In other words, these galaxies are not distinguishable from any other GOODS-ALMA galaxies based on their SED and thus recent SFH.
Therefore, despite their different properties (compactness, low depletion time) these sources manage to regulate themselves and stay within the MS.
They maintain their star-formation activity despite their lack of gas.
The presence of these sources and their ability to self-regulate and stay within the MS show a more complex picture of the MS than previously thought.

\section{\label{conclusions}Conclusions}

In this work, we aimed at constraining the recent SFH of the GOODS-ALMA galaxy sample presented in \cite{GomezGuijarro22a,GomezGuijarro22b} to put constraints on the different evolutionary scenarios they proposed to explain the presence of galaxies with starburst properties lying on the MS.

To conduct this analyse, we included in \cigale\ non-parametric SFHs.
Based on simulated SFHs, we showed that these models provide a better estimate of the stellar mass of galaxies than parametric SFH as well as more accurate SFR. 
These results confirmed the conclusions reached by \cite{Lower20} and \cite{Leja21}.
In this work, we use a ``bursty'' continuity prior, as introduced by \cite{Tacchella21a}.
We showed that the non parametric SFH is sensitive to strong starburst variations and rapid quenching events, with a slight overestimate of the SFR in case of strong quenching occurring less than 100\,Myr.
To characterise the SFH, we define the SFR gradient ($\nabla$SFR) which indicates the trend of the SFH over a given time by computing the angle between the $\Delta$SFR and $\Delta$M$_*$ variations over this given time.
A $\nabla$SFR close to 0$\degre$ is obtained for constant SFH while a value close to $+$($-$)90$\degre$ indicates a strong starburst (rapid quenching) event.

We selected the 65 GOODS-ALMA galaxies detected with \textit{Herschel} that are not optically dark.
These galaxies are fitted with \cigale\ using the non-parametric SFHs.
GOODS-ALMA galaxies have a $\nabla$SFR$_{100}$ ranging from -50$\degre$ to 75$\degre$.
The wide distribution of measured $\nabla$SFR$_{100}$ translates the stocchasticity of the recent SFH while the long-term SFH can be observed from $\nabla$SFR$_{1000}$.

The ``SB in the MS'' have positive to weak negative SFR gradients over the last 100\,Myr which is not compatible with an evolutionary path where these galaxies could come from the SB region above the MS. 
One would have expected strong negative gradients, indicative of a rapid quenching due to gas exhaustion in this case.
This results holds even when considering long timescales (300\,Myr and 1\,Gyr), confirming the absence of rapid quenching over the last Gyr.
Therefore, our results rule out the second scenario proposed in \cite{GomezGuijarro22b} where the ``SB in the MS'' could be galaxies transitioning from the SB region to stay on the MS or all through quiescence.

We find no differences in the last 1\,Gyr SFH between the galaxies classified as starburst in the MS by \cite{GomezGuijarro22b} and the other GOODS-ALMA galaxies.
Despite their different properties (compactness, low depletion time) the ``SB in the MS" have similar SFH than any other GOODS-ALMA galaxies and can not be identified from their SED, hence SFH.
In other words, these galaxies manage to maintain a star-formation activity allowing them to stay within the MS, like a self-regulation, despite their lack of gas.
The particularities of this sub-population highlight a diversity within the MS resulting in a more complex view of the relation.


\begin{acknowledgements}
L.C. warmly thanks M.~Boquien and Y.~Roehlly for their help in implementing the non-parametric SFHs in \cigale\ and O.~Ilbert for useful discussions.
This project has received financial support from the CNRS through the MITI interdisciplinary programs.
S.J. acknowledges the Villum Fonden research grants 37440, 13160 and the financial support from European Union's Horizon research and innovation program under the Marie Sk\l{}odowska-Curie grant agreement No. 101060888.
M.F. acknowledges NSF grant AST-2009577 and NASA JWST GO Program 1727.
H.I. acknowledges support from JSPS KAKENHI Grant Number JP19K23462.
G.E.M. acknowledges the Villum Fonden research grants 13160 and 37440 and the Cosmic Dawn Center of Excellence funded by the Danish National Research Foundation under the grant No. 140.
\end{acknowledgements}

\bibliographystyle{aa}
\bibliography{paper_goods_alma_sfh}
\end{document}